\documentclass
[showkeys,showpacs,superscriptaddress,preprintnumbers,byrevtex]{revtex4}
\usepackage{graphicx}
\usepackage{bm}
\begin{document}
\preprint{KIAS-P11050, CYCU-HEP-11-16}
\title{QCD Chiral restoration at finite $T$ under the Magnetic field:\\
Studies based on the instanton vacuum model\\}
\author{Chung-Wen Kao}
\email[E-mail: ]{cwkao@cycu.edu.tw}
\affiliation{Department of Physics, Chung-Yuan Christian University, Chung-Li 32023, Taiwan}
\author{Seung-il Nam}
\email[E-mail: ]{sinam@kias.re.kr}
\affiliation{School of Physics, Korea Institute for Advanced Study (KIAS), Seoul 130-722, Republic of Korea}
\date{\today}
\begin{abstract}
We investigate the chiral restoration at finite temperature $(T)$ under the strong external magnetic field $\vec{B}=B_{0}\hat{z}$ of
the SU(2) light-flavor QCD matter.
We employ the instanton-liquid QCD vacuum configuration accompanied with the linear Schwinger method for inducing the magnetic field. The Harrington-Shepard caloron solution is used to modify the instanton parameters, i.e. the average instanton size $(\bar{\rho})$ and inter-instanton distance $(\bar{R})$, as functions of $T$.
In addition, we include the meson-loop corrections (MLC) as the large-$N_{c}$ corrections because they are critical for reproducing the universal chiral restoration pattern. We present the numerical results for the constituent-quark mass as well as chiral condensate which signal the spontaneous breakdown of chiral-symmetry (SB$\chi$S), as functions of $T$ and $B$.
Besides we find that the changes for the $F_\pi$ and $m_\pi$ due to the magnetic field is relatively small, in comparison to those caused by the finite $T$ effect.
\end{abstract}
\maketitle

\section{Introduction}
Studies on the breakdown of symmetries and their restorations have been very useful in the analysis of phenomena related to phase transitions. Such studies applied to Quantum Chromodynamics are particularly fascinating since QCD owns a very complicated phase structure. Among them, the restoration of chiral symmetry at finite temperature ($T$) and/or quark chemical density ($\mu$) has been one of the most interesting and stimulating subjects for decades.

Recently it has been reported that very strong magnetic field in the order of the several times of $m^{2}_{\pi}\,[\mathrm{GeV}^{2}]$ can be produced in the noncentral (peripheral) heavy-ion collision (HIC) experiment by STAR collaboration at RHIC~\cite{:2009txa,:2009uh}. According to this strong magnetic field and $CP$-violating domains created inside the QGP, signals for possible $P$ and $CP$ violations were observed as the charge separation along the direction of the external magnetic field, which is perpendicular to the collision plane. The charge separation at relatively low $T$ has been already investigated within the instanton-vacuum framework by one of the authors (S.i.N.)~\cite{Nam:2009jb,Nam:2010nk}, in which related works and references can be found. Actually even before this sort of studies receiving much interest, QCD under magnetic field had been an important subject because of magnetar. There have been many approaches to study this issue.
Here we will present our results based on the instanton-vacuum model \cite{Nam:2010mh,Nam2011}.

\section{The instanton vacuum model with MLC at finite T and B}
 We employ the  instanton-liquid model at finite $T$, the four-dimensional integral in the effective action is replaced by the three-dimensional integral with the summation over the fermionic Matsubara frequency. The Harrington-Shepard caloron solution ~\cite{Harrington:1976dj}is used to modify the instanton parameters, i.e. the average instanton size $(\bar{\rho})$ and inter-instanton distance $(\bar{R})$, as functions of $T$. The result is shown in Fig.(1). Although this approach does not manifest the quark confinement, such as the nontrivial holonomy caloron, i.e. the Kraan-van Baal-Lee-Lu caloron~\cite{Kraan:1998pm,Lee:1998bb},  as shown in Refs.~\cite{Nam:2009jb,Nam:2010nk,Nam:2009nn}, it is a useful nonperturbatve method to study QCD matter at finite $T$. To include the external magnetic field, we make use of the linear Schwinger method~\cite{Schwinger:1951nm}. Besides we also take into account the meson-loop corrections (MLC) for the SU(2) light-flavor sector as the large-$N_c$ corrections because it is essential to reproduce the correct current-quark mass dependence of relevant physical quantities~\cite{Nam:2008bq} and universal-class chiral restoration pattern at finite $T$~\cite{Nam:2010mh}. The detail of our model we refer readers to our papers \cite{Nam:2010mh,Nam2011}.

\begin{figure}
\centering
\includegraphics[width=4.0cm]{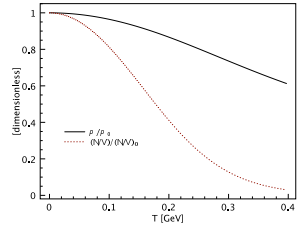}
\caption{Normalized $\bar{\rho}/\bar{\rho}_{0}$ and ${\cal N}/{\cal N}_{0}$ as a function of $T$ for $N_{c}=3$, where $\mathcal{N}\equiv N/V$}
\label{FIG1}
\end{figure}

\section{Our results}
Our results of the constituent quark masses are shown in
Fig.(2) and Fig.(3) for $T$=0 and $T$=50 NeV. The chiral condensates are shown in Fig.(4) for the chiral limit and physical quakr mass cases, respectively.
The critical $T$ for the chiral restoration, $T_c$ tends to be shifted higher pronouncedly in the presence of the $B_0$ in the chiral limit as shown in Table.(1).
The strength of the isospin breaking between the $u$ and $d$ quark condensates is also explored in detail by defining the ratio $\mathcal{R}\equiv(\langle iu^\dagger u\rangle-\langle id^\dagger d\rangle)/(\langle iu^\dagger u\rangle+\langle id^\dagger d\rangle)$ as a function of $T$ and $B_0$. They are shown in Fig.(5).
Finally, we compute the pion weak-decay constant $F_\pi$ and pion mass $m_\pi$ below $T_c$ as functions of $T$ and $B_0$ shown in Fig.(6).

\begin{figure}
\centering
\begin{tabular}{cc}
\includegraphics[width=5.0cm]{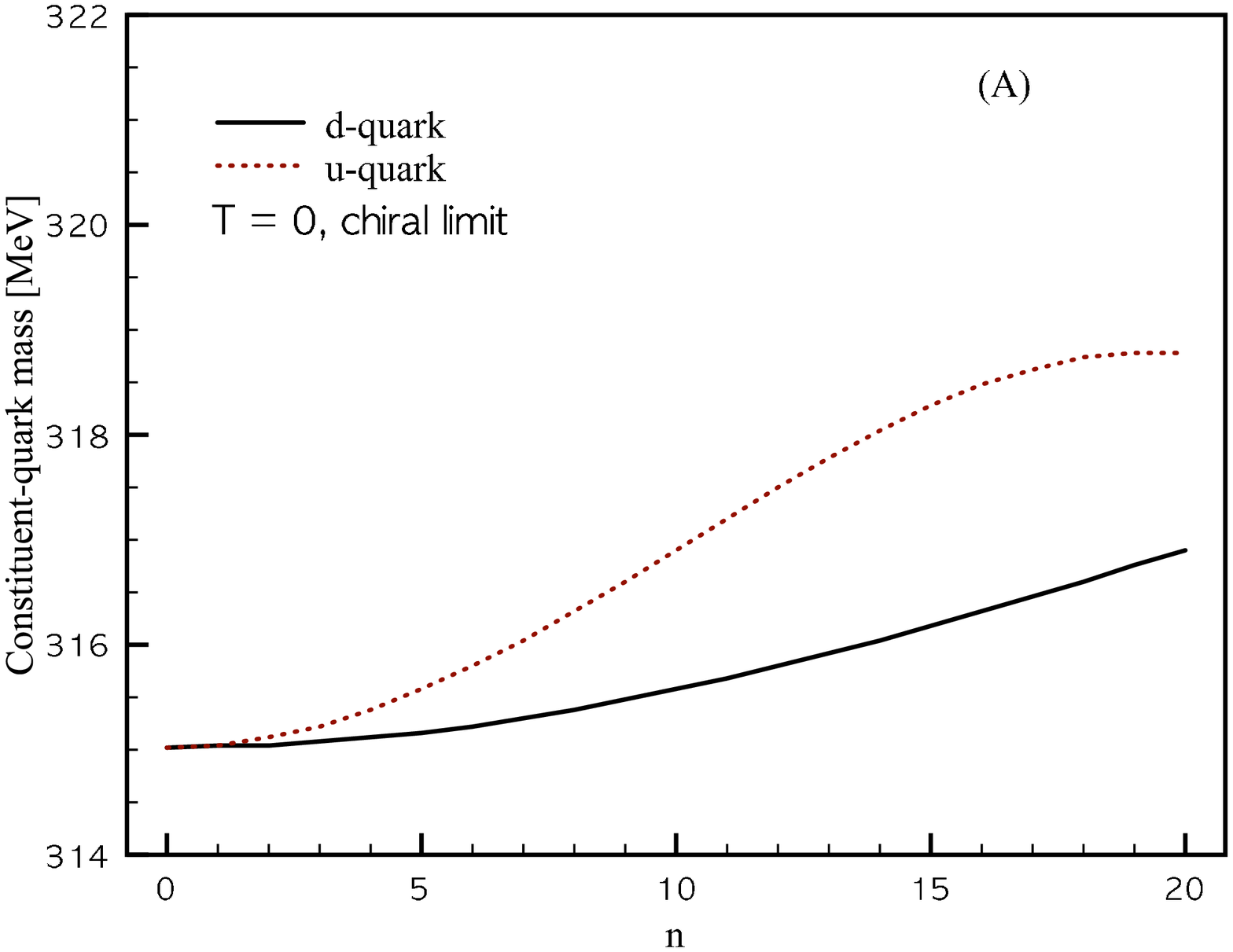}
\includegraphics[width=5.0cm]{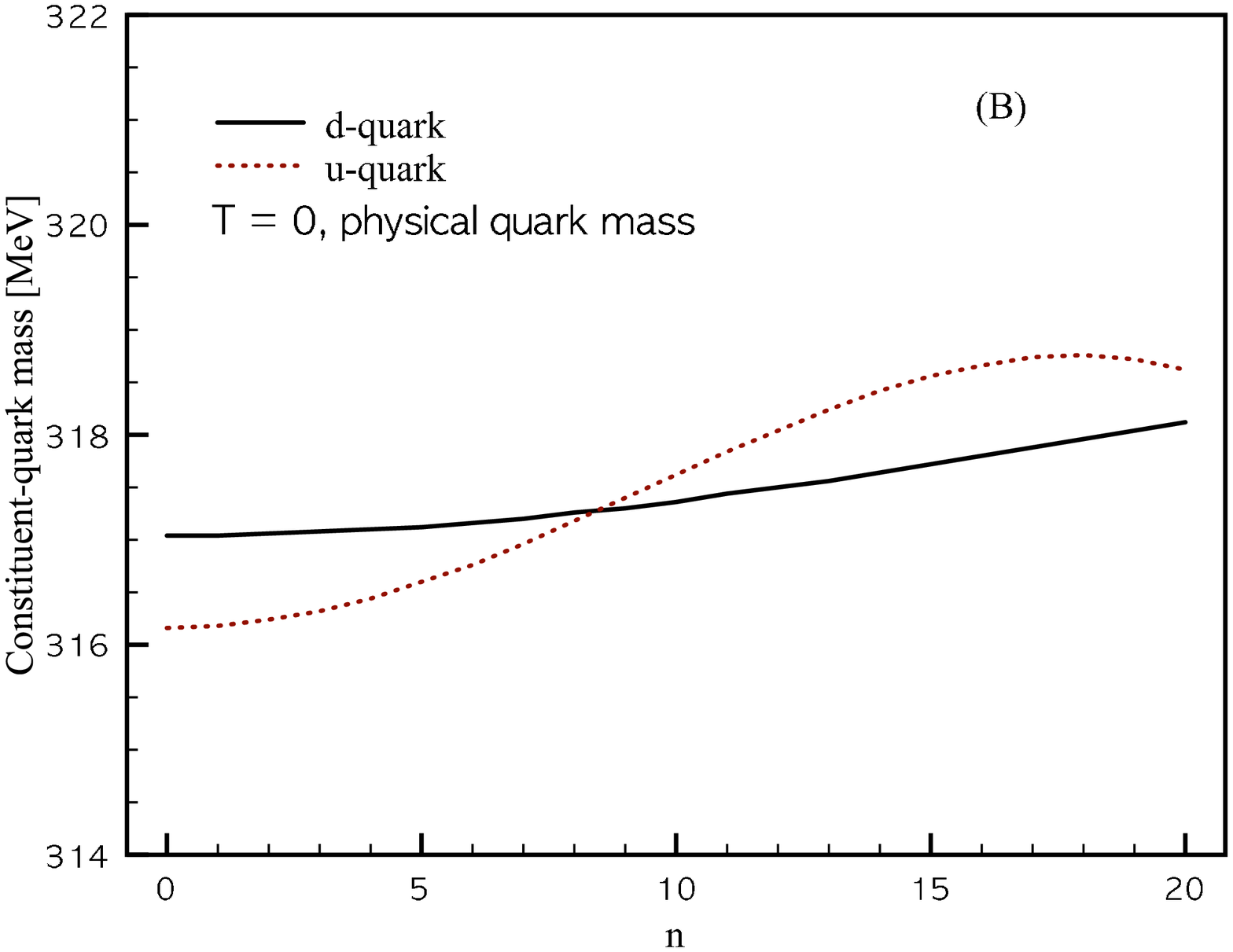}
\end{tabular}
\caption{Constituent-quark mass ($M_f$) at $T=0$ as a function of $n\equiv eB_0/m^2_\pi$ for the chiral limit (left column) and physical quark mass (right column).}
\end{figure}

\begin{figure}
\centering
\begin{tabular}{cc}
\includegraphics[width=5.0cm]{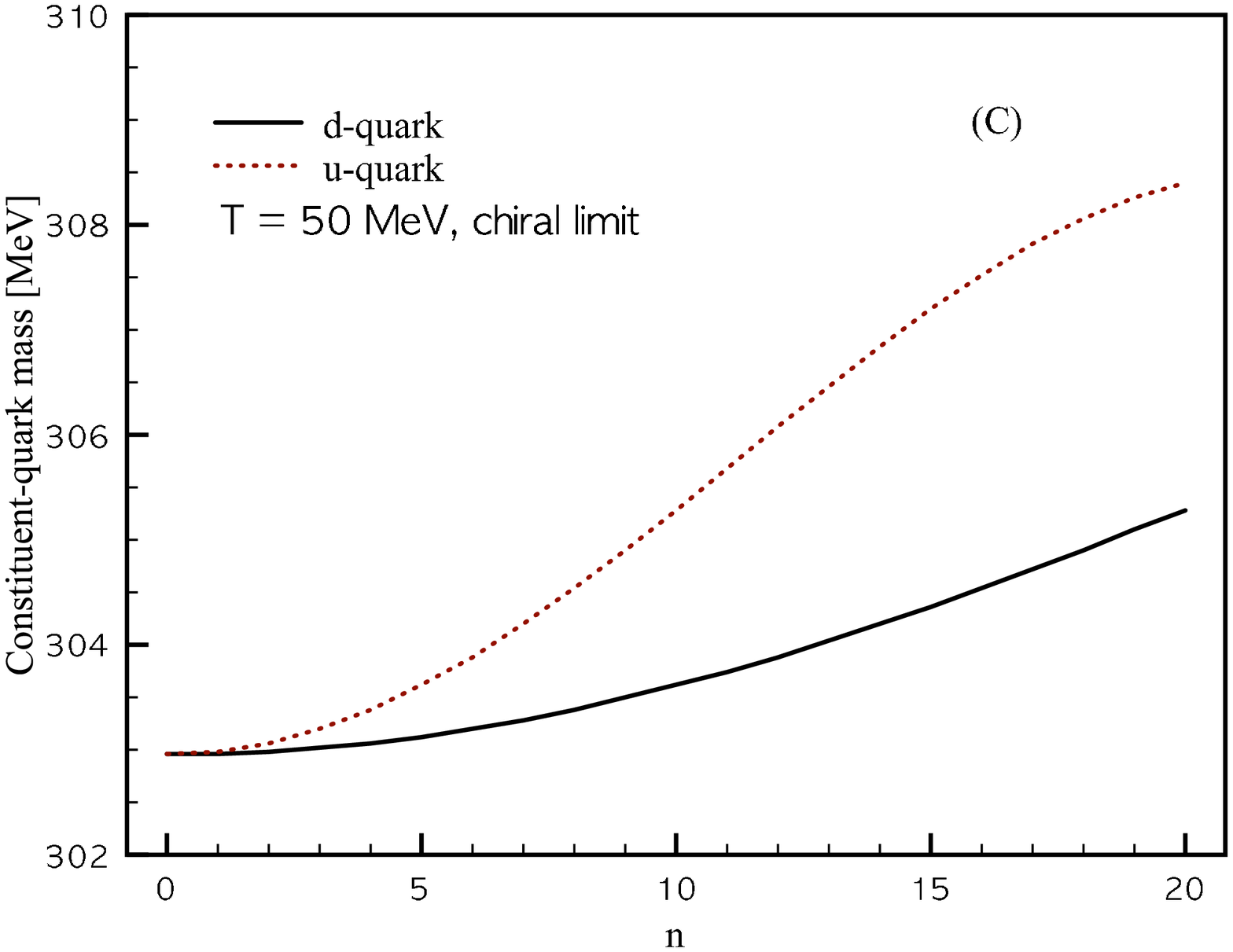}
\includegraphics[width=5.0cm]{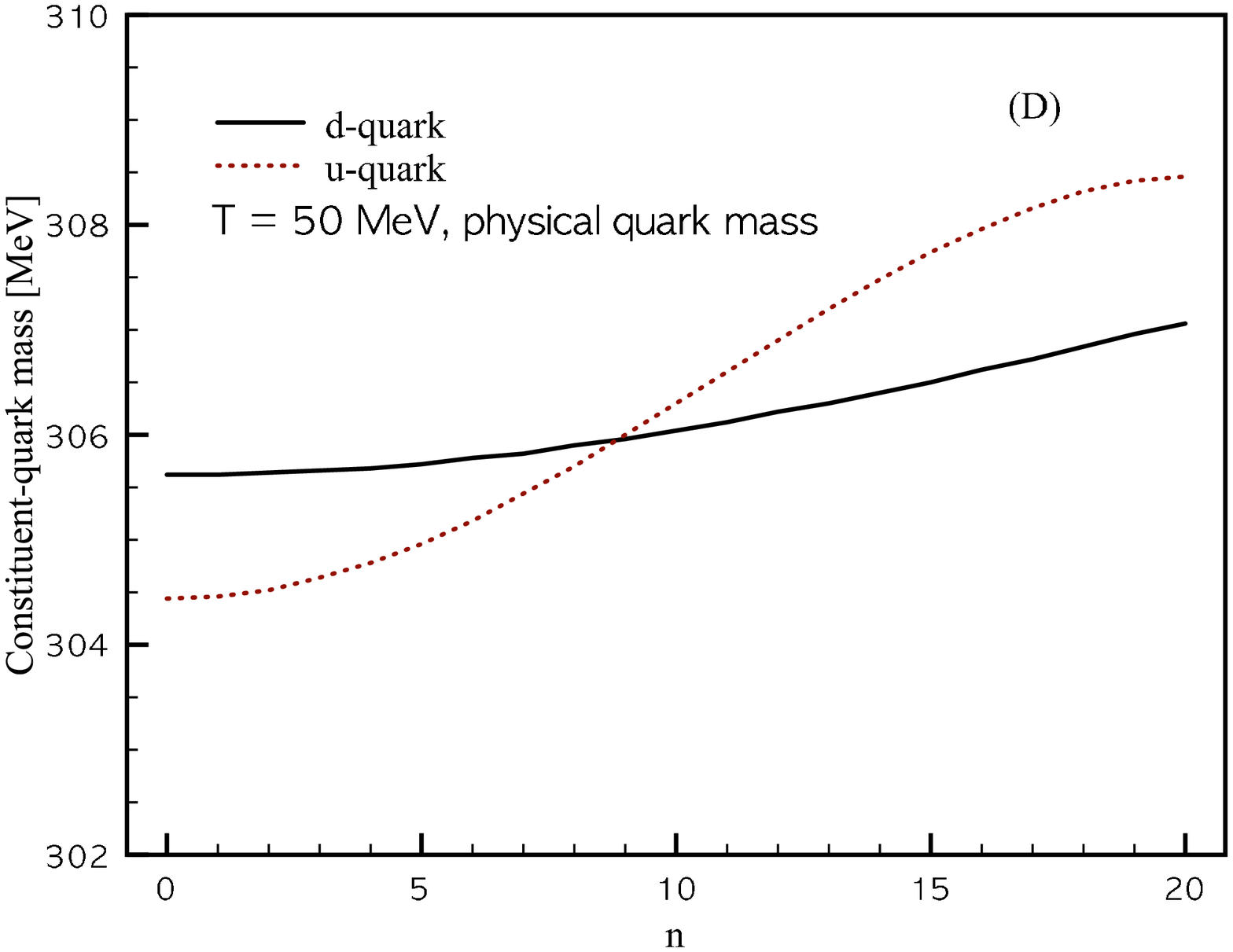}
\end{tabular}
\caption{Constituent-quark mass ($M_f$) $T=50$MeV as a function of $n\equiv eB_0/m^2_\pi$ for the chiral limit (left column) and physical quark mass (right column).}
\end{figure}

\begin{figure}[t]
\centering
\begin{tabular}{cc}
\includegraphics[width=5.0cm]{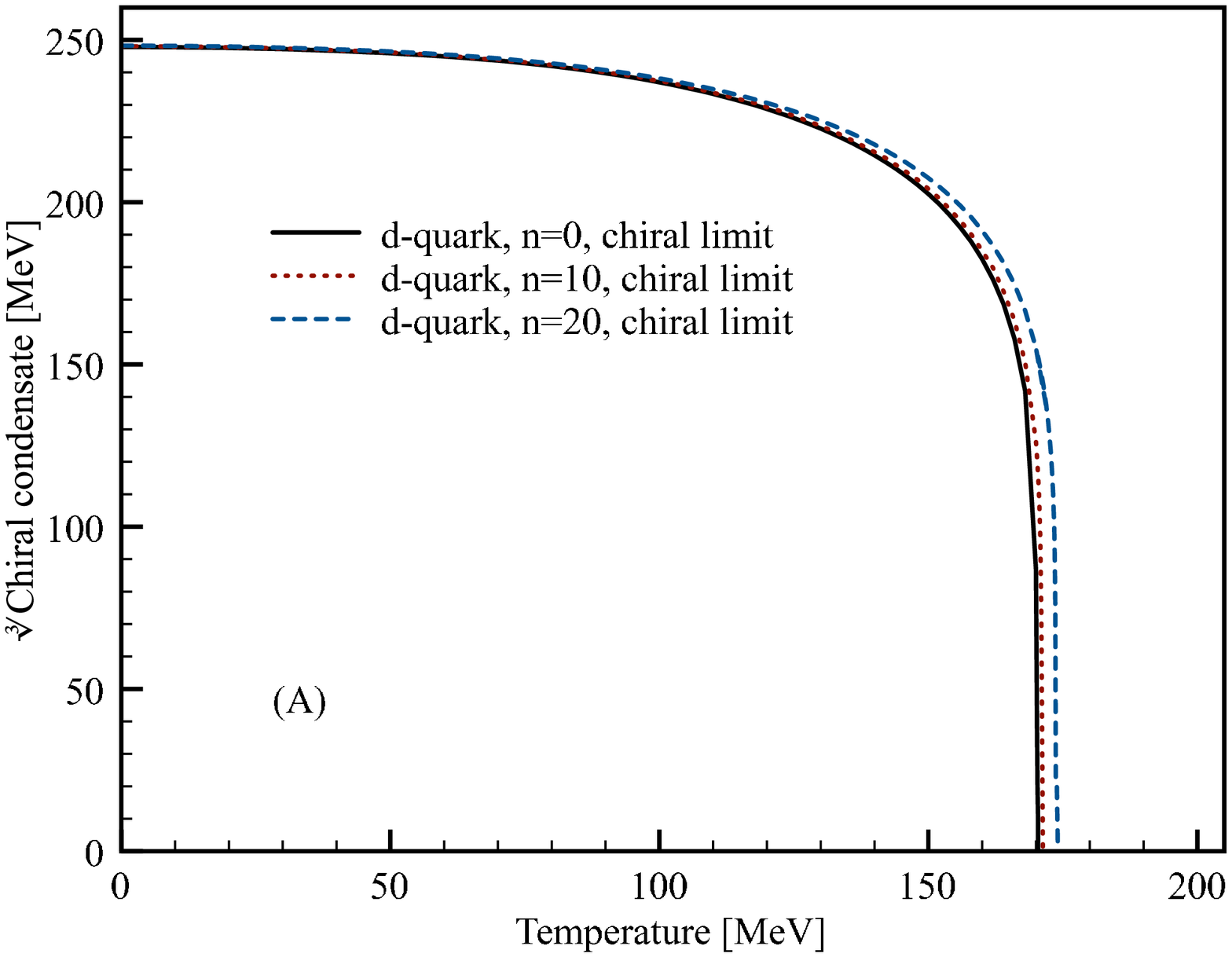}
\includegraphics[width=5.0cm]{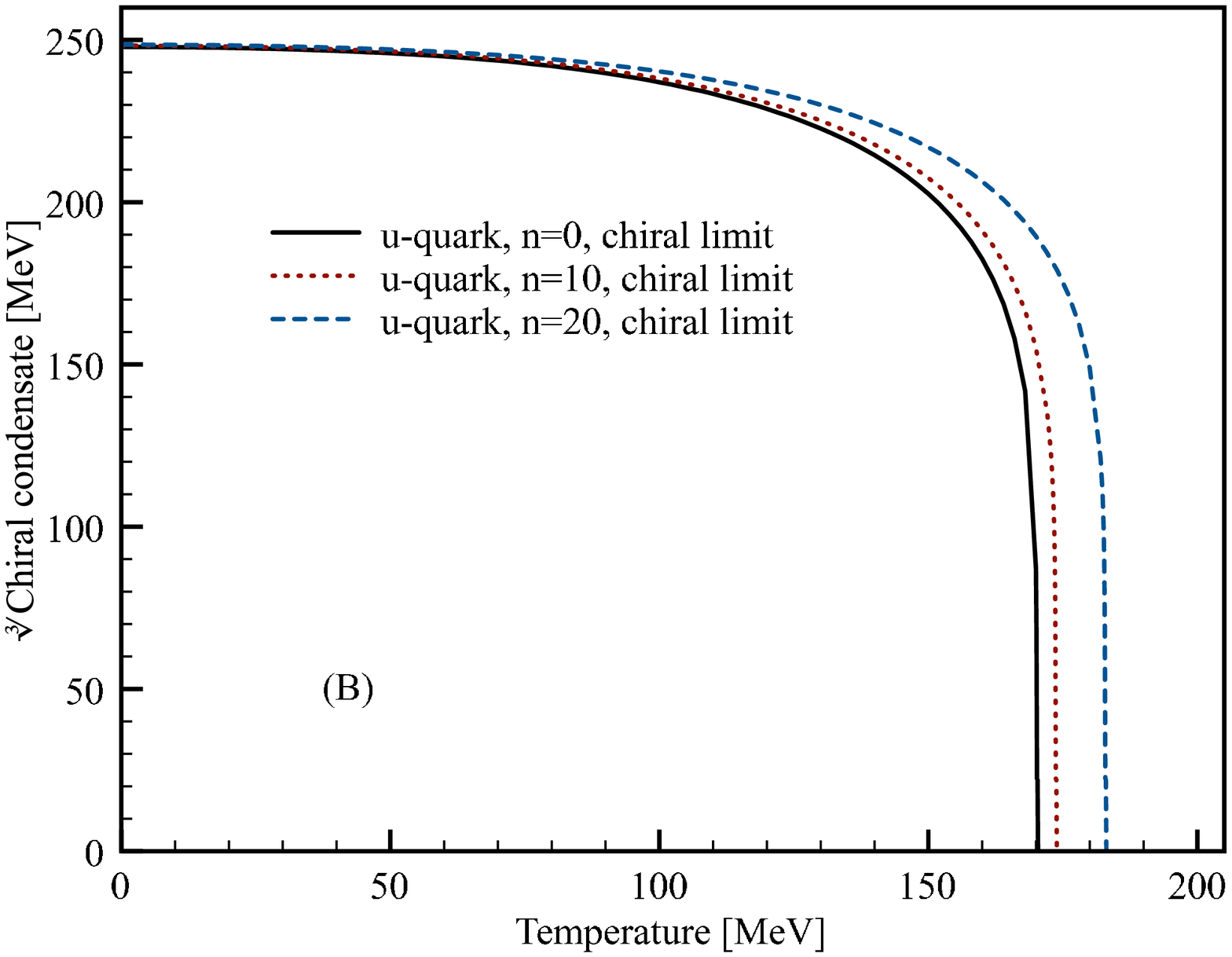}
\end{tabular}

\begin{tabular}{cc}
\centering
\includegraphics[width=5.0cm]{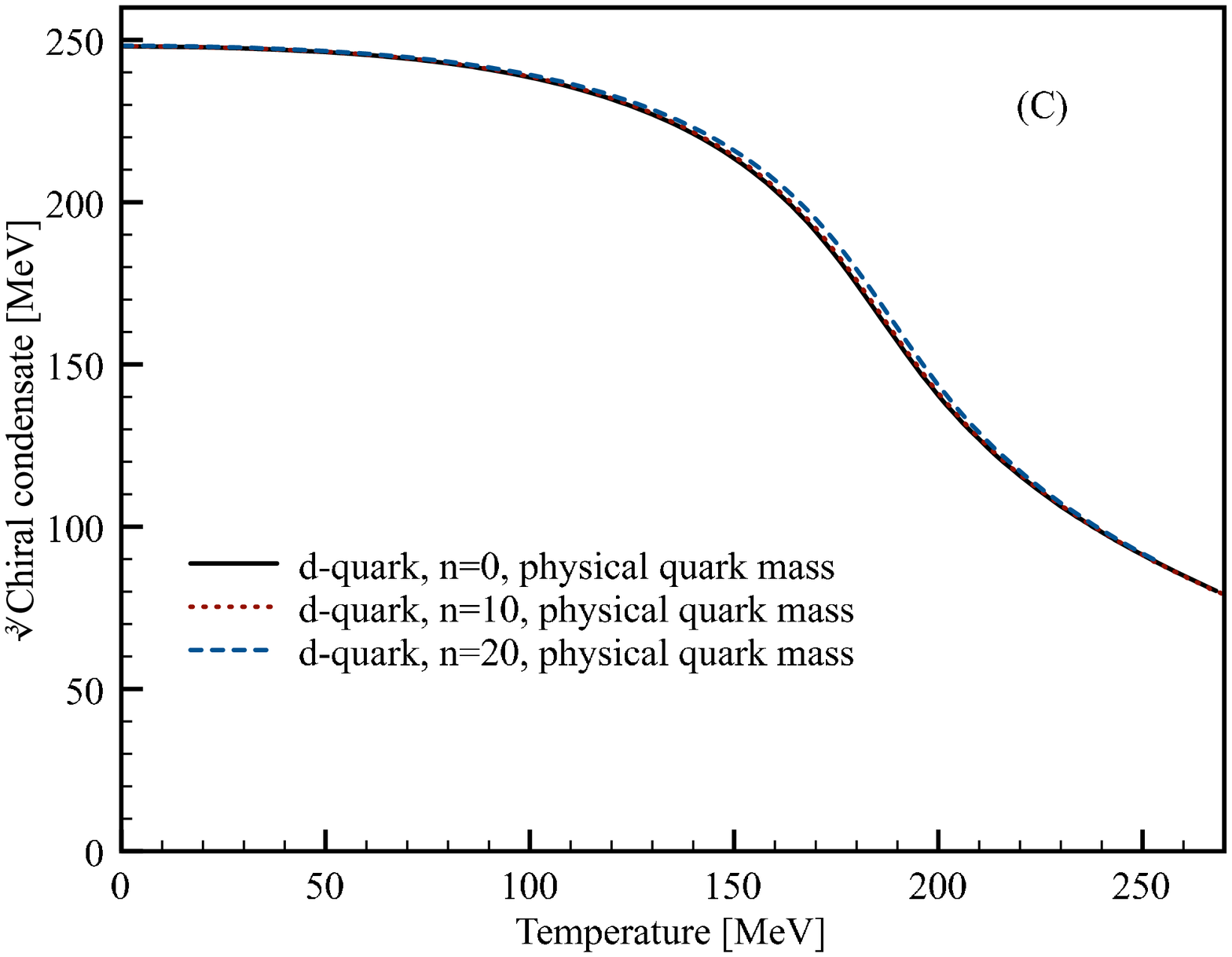}
\includegraphics[width=5.0cm]{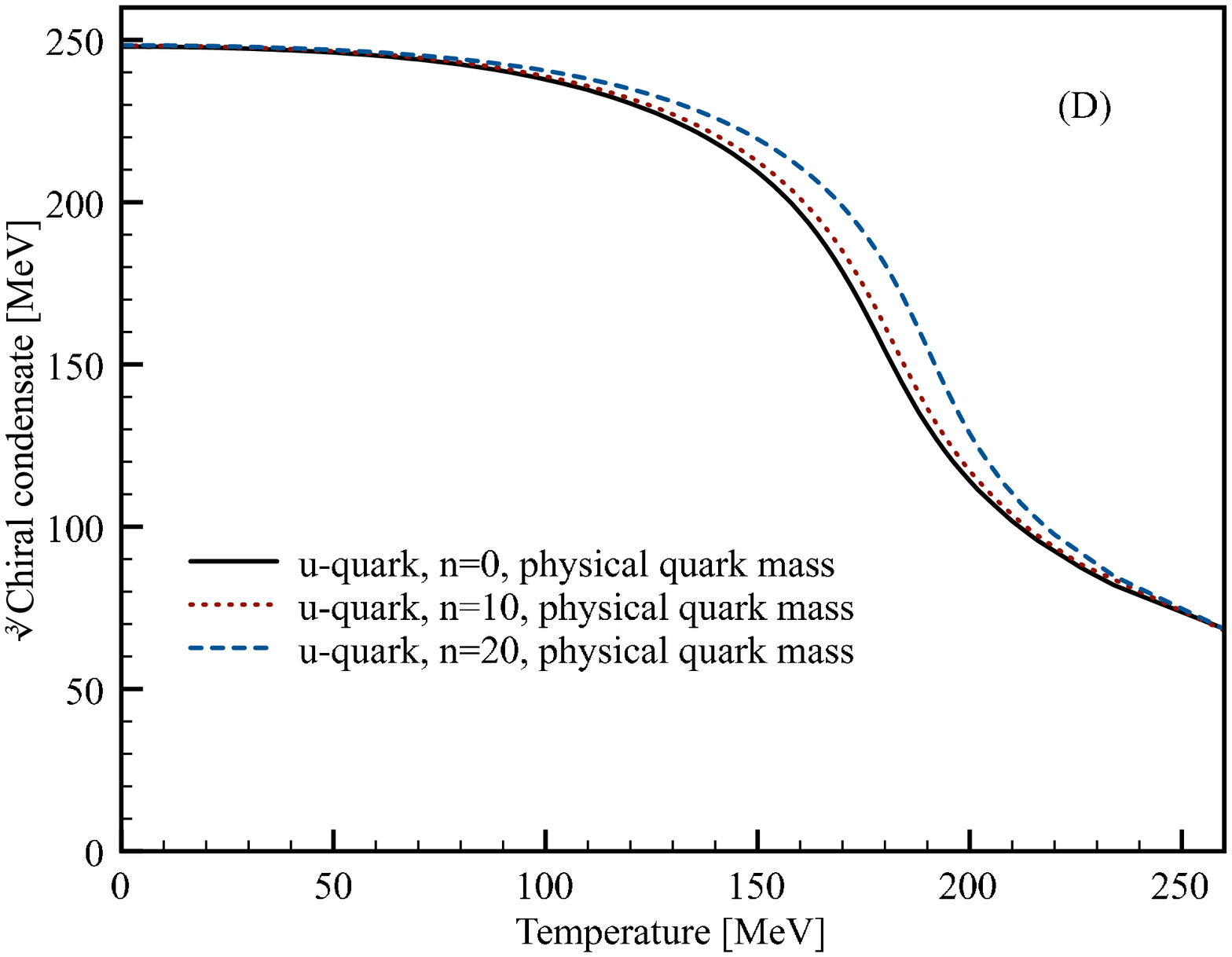}
\end{tabular}
\caption{Chiral condensate as a function of $T$. The numerical results for the chiral limit and physical quark mass are shown in the left and right columns. The results for the $d$ and $u$ quarks are given in the upper and lower rows, respectively.}
\label{FIG3}
\end{figure}

\begin{figure}
\centering
\begin{tabular}{cc}
\includegraphics[width=4.2cm]{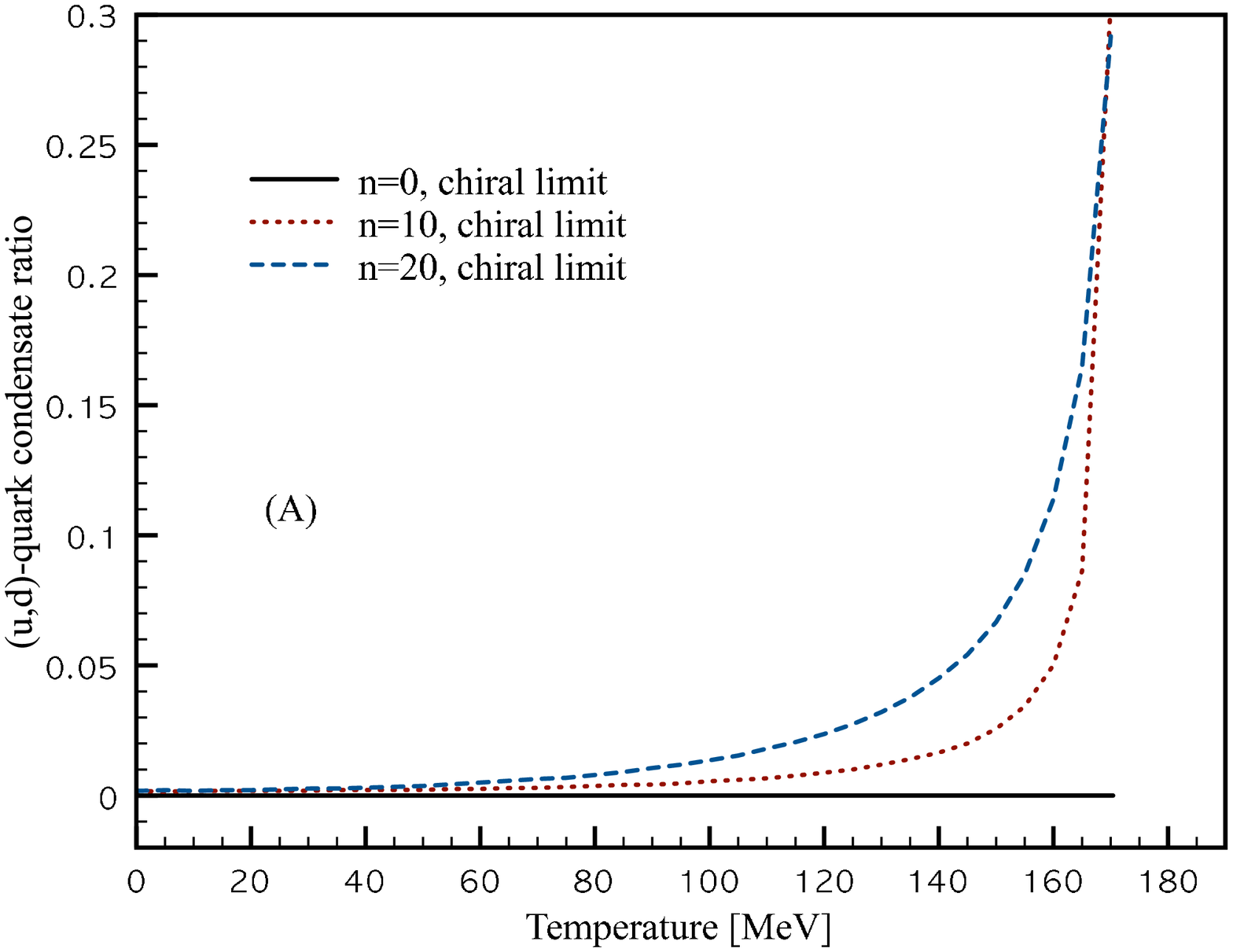}
\includegraphics[width=4.2cm]{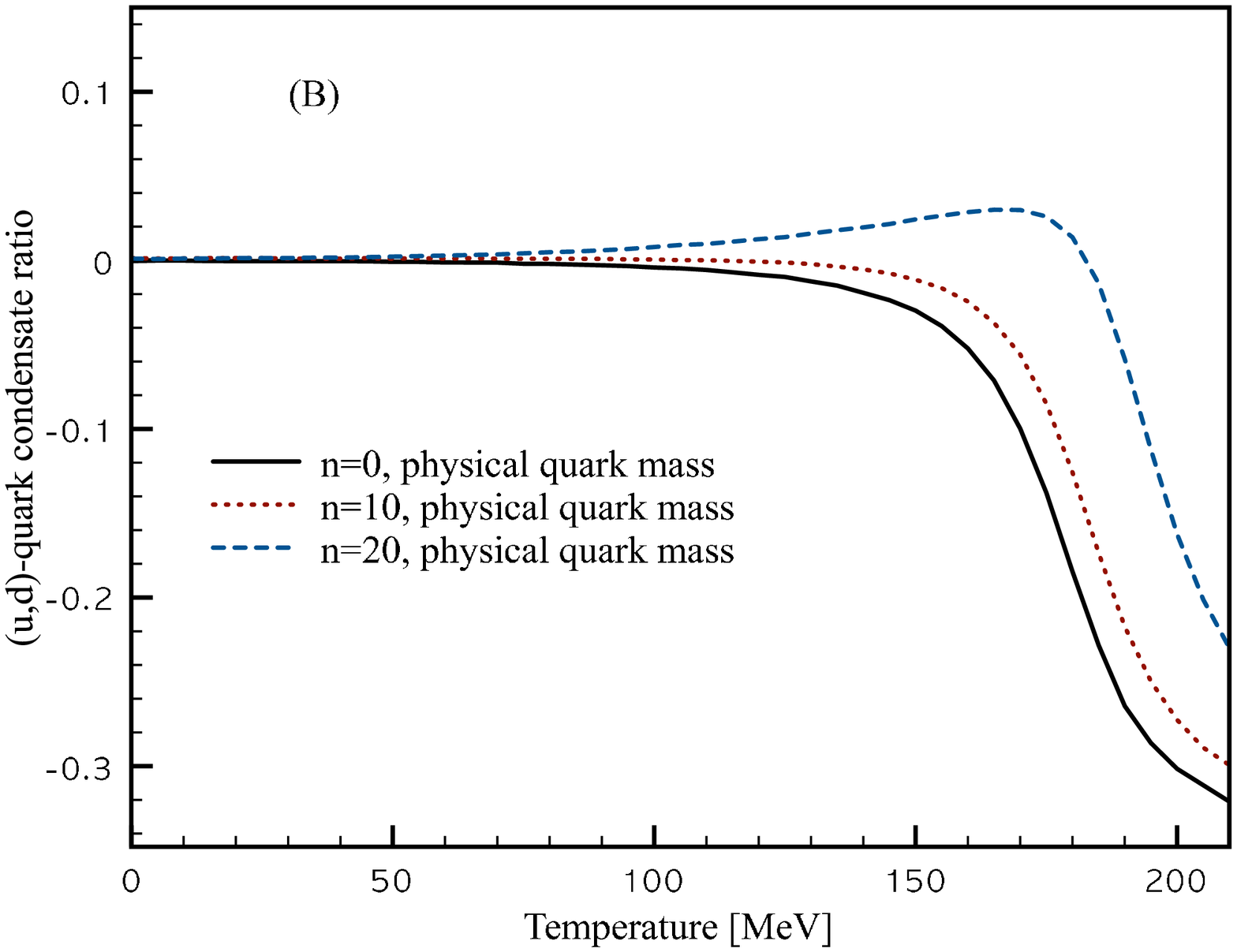}
\end{tabular}
\caption{$(u,d)$-quark condensate ratio $\mathcal{R}$ as function of $T$ for different $n\equiv eB_0/m^2_\pi$ values. The numerical results for the chiral limit and physical quark mass are shown in the left and right panels, respectively.}
\label{FIG4}
\end{figure}

\begin{figure}[t]
\centering
\begin{tabular}{cc}
\includegraphics[width=4.2cm]{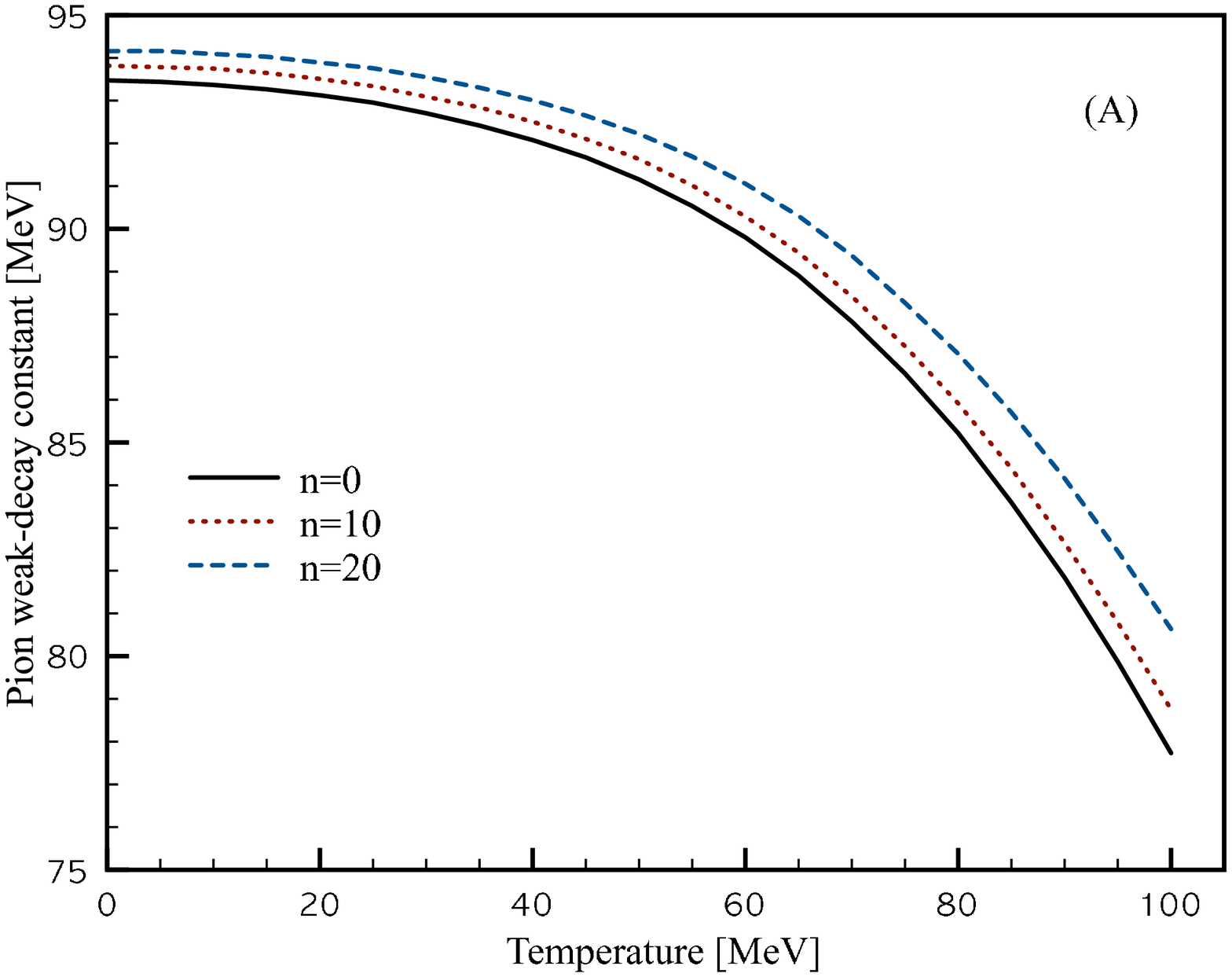}
\includegraphics[width=4.2cm]{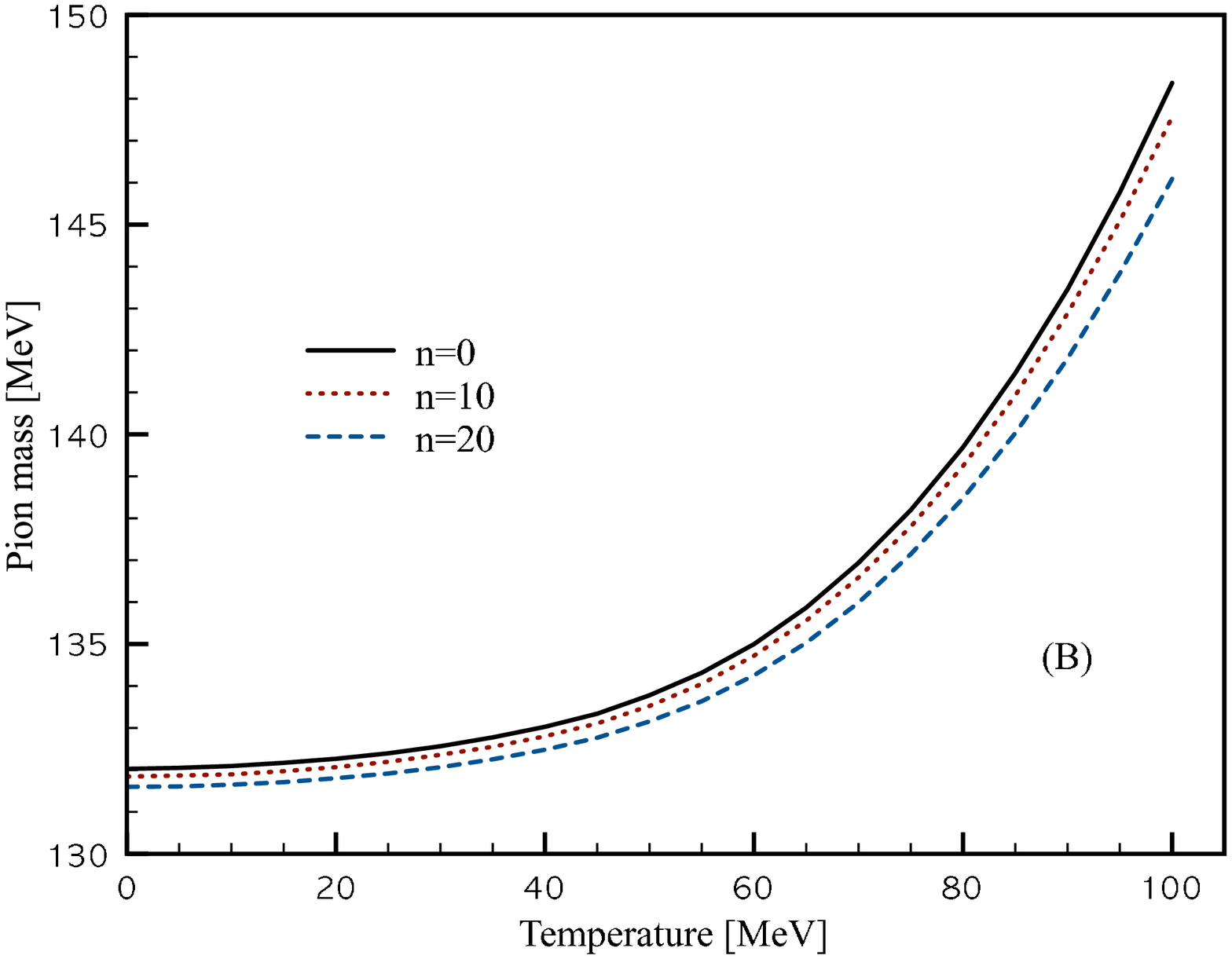}
\end{tabular}
\caption{Pion weak-decay constant $F_\pi$ (A) and pion mass $m_\pi$ (B) as a function of $T$, varying the strength of the magnetic field.}
\end{figure}

\section{Conclusions and Outlooks}

\begin{table}
\centering
\begin{tabular}{c|c|c|c}
&$n=0$&$n=10$&$n=20$\\
\hline\noalign{\smallskip}
$T^{u}_c$&$170.4$ MeV&$173.9$ MeV&$183.1$ MeV\\
$T^{d}_c$&$170.4$ MeV&$171.3$ MeV&$174.1$ MeV\\
\end{tabular}
\caption{Critical temperature for the $u$ and $d$ flavors, $T^{u}_c$ and $T^{d}_c$, for the chiral limit in the presence of the external magnetic field, $n=eB_0/m^2_\pi$.}
\label{TABLE1}
\end{table}

We summarize our results as follows,
\begin{itemize}
\item  The external magnetic field enhances the SB$\chi$S in terms of the magnetic catalysis \cite {Miransky:2002rp}, which is proportional to $(e_fB_0)^2$. Hence, the $u$-quark constituent mass is more sensitive to the magnetic field.
\item   On top of the explicit isospin symmetry breaking, there appears a point at which the constituent quark masses for the $u$ and $d$ quarks coincide each other for the strong magnetic field $eB_0\approx10^{19}$ G.
\item
    The effects from the magnetic catalysis becomes more pronounced in the higher $T$ region because as $T$ increases the SB$\chi$S effects generated from the instanton is weakened. Thus the magnetic catalysis effects becomes more important there.
\item For the physical quark mass case, the ratio $\mathcal{R}$ shows nontrivial structures with respect to $T$ and $B_{0}$ due to the complicated competition between the magnetic catalysis and the explicitly isospin breaking effect.
\item According to our simple and qualitative analysis using the GOR relation, we observe correct partial chiral-restoration and magnetic-catalysis behaviors for the pion-weak decay constant $F_\pi$ and pion mass $m_\pi$. They decreases and increases about $10\%$ at $T\approx100$ MeV in comparison to those at $T=0$, respectively. However, the changes due to the magnetic field are relatively small, just a few percent.
\end{itemize}
Our next step is to include the finite $\mu$ to the effective action in the instanton framework. The result will be reported later.

\section*{Acknowledgements}
The work of S.i.N. was supported by the grant NRF-2010-0013279 from National Research Foundation (NRF) of Korea. The work of C.W.K. was supported by the grant NSC 99-2112-M-033-004-MY3 from National Science Council (NSC) of Taiwan. C.W.K has also acknowledged the support of NCTS (North) in Taiwan.


\begin{thebibliography}{99}
\bibitem{:2009txa}
  B.~I.~Abelev {\it et al.}  [STAR Collaboration],
  Phys.\ Rev.\  C {\bf 81}, 054908 (2010).
\bibitem{:2009uh}
  B.~I.~Abelev {\it et al.}  [STAR Collaboration],
  Phys.\ Rev.\ Lett.\  {\bf 103}, 251601 (2009).
\bibitem{Nam:2009jb}
  S.~i.~Nam,
  Phys.\ Rev.\  D {\bf 80}, 114025 (2009).
\bibitem{Nam:2010nk}
  S.~i.~Nam,
  Phys.\ Rev.\  D {\bf 82}, 045017 (2010).
\bibitem{Nam:2010mh}
  S.~i.~Nam and C.~W.~Kao,
  Phys.\ Rev.\  D {\bf 82}, 096001 (2010).
\bibitem{Nam2011}S.~i.~Nam and C.~W.~Kao,
  Phys.\ Rev.\  D {\bf 83}, 096009 (2011).
\bibitem{Harrington:1976dj}
  B.~J.~Harrington and H.~K.~Shepard,
  Nucl.\ Phys.\  B {\bf 124}, 409 (1977).
\bibitem{Kraan:1998pm}
  T.~C.~Kraan and P.~van Baal,
  Nucl.\ Phys.\  B {\bf 533}, 627 (1998).
\bibitem{Lee:1998bb}
  K.~M.~Lee and C.~h.~Lu,
  Phys.\ Rev.\  D {\bf 58}, 025011 (1998)
\bibitem{Nam:2009nn}
  S.~i.~Nam,
  J.\ Phys.\ G {\bf 37}, 075002 (2010).
\bibitem{Schwinger:1951nm}
  J.~S.~Schwinger,
  Phys.\ Rev.\  {\bf 82}, 664 (1951).
\bibitem{Nam:2008bq}
  S.~i.~Nam,
  Phys.\ Rev.\  D {\bf 79}, 014008 (2009).
\bibitem{Miransky:2002rp}
  V.~A.~Miransky and I.~A.~Shovkovy,
  Phys.\ Rev.\  D {\bf 66}, 045006 (2002).
\end{thebibliography}


\end{document}